\documentclass[%
superscriptaddress,
nofootinbib,
prd
]{revtex4-2}
\usepackage{float}
\usepackage[normalem]{ulem}
\usepackage{graphicx}% Include figure files
\usepackage{dcolumn}% Align table columns on decimal point
\usepackage{bm}% bold math
\usepackage{amsmath}
\usepackage{amsfonts}
\usepackage{siunitx}
\usepackage[english]{babel}
\usepackage{amsfonts}
\usepackage{braket,slashed,bm}
\usepackage{natbib}
\usepackage[dvipsnames]{xcolor}
\usepackage[caption=false,labelformat=simple]{subfig}

\usepackage{multirow}

\usepackage{hyperref}
\hypersetup{
    colorlinks=true,
    linkcolor=blue,
    filecolor=blue,      
    urlcolor=blue,
    anchorcolor=blue,
    citecolor=blue
}
\def\beq{\begin{equation}}
\def\eeq{\end{equation}}
\def\barr{\begin{array}}
\def\earr{\end{array}}
\def\dis{\displaystyle}

\newcommand{\be}{\begin{equation}}
\newcommand{\ee}{\end{equation}}
\newcommand{\bea}{\begin{eqnarray}}
\newcommand{\eea}{\end{eqnarray}}

\newcommand{\bi}{\begin{itemize}}
	\newcommand{\ei}{\end{itemize}}

\begin{document}

\preprint{APS/123-QED}

\title{Tri-bimaximal-Cabibbo Mixing: Flavour violations in the charged lepton sector}
\author{Mathew Thomas Arun}%
\email{thomas.mathewarun@gmail.com}
\author{Anirudhan A. Madathil}%
\email{anirudhanam777@gmail.com}
\affiliation{School of Physics, Indian Institute of Science Education and Research Thiruvananthapuram, Vithura, Kerala, 695551, India}%
\date{\today}% It is always \today, today,

\begin{abstract}

The well understood structure of $U_{pmns}$ matrix mandates a Cabibbo mixing matrix in the first two generations of the charged lepton sector if we assume Tri-bimaximal mixing in the neutrino sector. This ansatz, called Tri-bimaximal-Cabibbo mixing, is ruled out immediately by the experiments searching for charged lepton flavour violating currents. In this article, we aim to show that the resurrection of the theoretically well motivated Tri-bimaximal mixing scenario comes naturally within Minimal Flavour Violation hypothesis in the lepton sector. We analyse the flavour violating currents $\mu\rightarrow e e e$, $\mu Ti \to e Ti$, $\mu \rightarrow e \gamma$, $\pi^0\rightarrow  e^+ \mu^{-}$ and $K_L \rightarrow  \mu^+ e^-$ in this scenario and show that the New Physics that generates mixing among the charged lepton could lie within the reach of hadron colliders. In the minimal field content scenario, though the most stringent constrain on New Physics is $\gtrsim \mathcal{O}(10$ TeV) for maximal coupling, considering more natural couplings relaxes it to $\gtrsim \mathcal{O}(4$ TeV). On the other hand, New Physics with the extended field content is even more strongly constrained to $\gtrsim \mathcal{O}(75$ TeV) for maximal coupling, while it gets relaxed to $\gtrsim \mathcal{O}(31$ TeV) for natural scenario.

\end{abstract}
\maketitle
\section{Introduction}

The discovery of neutrino oscillation \cite{Super-Kamiokande:1998kpq, *Super-Kamiokande:2001ljr, *SNO:2001kpb, *SNO:2002tuh} has been of fundamental importance in understanding the flavour mixing in the neutral lepton sector. One such mixing matrix, motivated from discrete family symmetry like the $A_{4}$, is the Tri-bimaximal mixing ($U_{TB}$)~\cite{Harrison:2002er}. But a non-zero measurement of the reactor angle by RENO \cite{RENO:2012mkc} and Daya Bay \cite{DayaBay:2012fng} experiments rule out this hypothesis. Nevertheless, since it is theoretically well motivated and the rest of the angles predicted match the experimental results well enough, various variants based on the Tri-bimaximal mixing have been postulated  \cite{King:2012vj,*King:2013eh,*King:2013xba,*Antusch:2012fb,*Antusch:2013ti,*deMedeirosVarzielas:2012cet,*Zhou:2012zj,*Jourjine:2012yu,*Ishimori:2012gv,*Ishimori:2012gv,*Bazzocchi:2012st,*Zhao:2014yaa,*Hu:2012eb}.

Tri-bimaximal-Cabibbo~\cite{King:2012vj} (TBC) mixing is one such ansatz that emerged from a Tri-bimaximal hypothesis by including the non-zero reactor angle in the mixing of charged lepton sector. But, mixing in the charged lepton sector will contribute to the lepton flavour violating processes like $\pi^{0} \rightarrow e^{+}\mu^{-}, K_{L} \rightarrow \mu^{+}e^{-}, \mu N \rightarrow e N, \mu\rightarrow e e e, \mu \rightarrow e \gamma$ etc. Such decays are searched for at high-intensity experiments giving rise to stringent bounds of BR$(\pi^{0} \rightarrow e^{+}\mu^{-}) < 3.2 \times 10^{-10}$ \cite{NA62:2021zxl} (NA62 CERN), BR$(K_{L} \rightarrow \mu^{+}e^{-}) < 4.7\times 10^{-12}$ \cite{BNL:1998apv} (BNL), BR$(\mu Ti\rightarrow e Ti)<6.1 \times 10^{-13}$ \cite{SINDRUMII:1998mwd} (SINDRUM II), BR$(\mu\rightarrow e e e)<1\times 10^{-12}$ \cite{SINDRUM:1987nra} (SINDRUM), BR$(\mu \rightarrow e \gamma) < 4.2 \times 10^{-13} $ (MEG) \cite{MEG:2016leq}. In this article, we discuss a model-independent approach of lepton flavour violation (LFV) induced by TBC mixing ansatz. Using the most general set of operators, TBC mixing without any flavour symmetry in the lepton sector is then ruled out by the current limits on LFV decays $\pi^{0} \rightarrow e^{+}\mu^{-}, K_{L} \rightarrow \mu^{+}e^{-}$ and $\mu \rightarrow e \gamma$.

 On the other hand, we show that the Minimal Flavour Violation (MFV) hypothesis~\cite{DAmbrosio:2002vsn,Cirigliano:2005ck}, with a choice of basis including the Cabibbo mixing among the charged leptons, can protect TBC mixing ansatz from these strong bounds. Under this hypothesis, all the operators in the lepton sector are assumed to transform under the flavour symmetry $SU(3)_{L}\times SU(3)_{e}$. We compute the limit on the lepton flavour violating New Physics (NP) scale, using dim-6 operators, for observables like $\mu\rightarrow e e e$, $\mu Ti \to e Ti$, $\mu \rightarrow e \gamma$, $\pi^0\rightarrow  e^+ \mu^{-}$ and $K_L \rightarrow  \mu^+ e^-$. We then extend the analysis to include right-handed neutrinos. Assuming modest values of couplings, we show that the NP, leading to the flavour violation in the former case, could lie within the LHC regime $\sim \mathcal{O}(4~\text{TeV})$. Whereas, in the extended field setup, the New Physics is strongly constrained to be much heavier $\sim \mathcal{O}(31~\text{TeV})$.

The article is organised as follows. In section:\ref{sec:tbclfv} we give a brief review of TBC mixing ansatz and compute the flavour violating leptonic decays of $K_L$, $\pi^0$ and $\mu \rightarrow e \gamma$ show that the current experimental bounds are violated, ruling out vanila scenario. We then discuss the TBC ansatz in MFV with minimal and extended field content in section:\ref{sec:MFVtbc}. The basis choice with the ansatz is then analysed in the decays of $\mu \to eee$, $\pi \to e \mu $, $\mu \rightarrow e \gamma, K_L \to \mu e$ and $\mu N \to e N$. In section:\ref{sec:conclusion} we give our remarks and results.

\section{TBC mixing induced LFV}
\label{sec:tbclfv}

In general the mixing in the lepton sector ($U_{pmns}$) can be written as, $U_{pmns} = {U^{L}_{e}}^{\dagger}U^{L}_{\nu}$. 
Where $U^{L}_{e}$ is the mixing in the charged lepton sector and $U^{L}_{\nu}$ is the mixing in the neutrino sector. If we assume the TBC ansatz then,

 \begin{equation}
     {U^{L}_{e}}^{\dagger} = 
     \begin{pmatrix}
      c^{e}_{12} & s^{e}_{12}e^{-i\delta^{e}_{12}} & 0\\
      -s^{e}_{12}e^{i\delta^{e}_{12}} & c^{e}_{12} & 0\\
      0 & 0 & 1\\
     \end{pmatrix};~~
     U^{L}_{\nu} =  U_{TB} =
     \begin{pmatrix}
     \sqrt{\frac{2}{3}} & \frac{1}{\sqrt{3}} & 0\\
     -\frac{1}{\sqrt{6}} &\frac{1}{\sqrt{3}} & \frac{1}{\sqrt{2}}\\
     \frac{1}{\sqrt{6}} &-\frac{1}{\sqrt{3}} & \frac{1}{\sqrt{2}}
     \end{pmatrix},
     \label{eq:ule}
 \end{equation}
where $c^{e}_{12} = \cos{\theta^{e}_{12}}$ and $s^{e}_{12}= \sin{\theta^{e}_{12}}$. Mixing in the first two generations of charged leptons gives a non-zero reactor angle in $U_{pmns}$, which satisfies the relation $\sin{\theta_{13}} = \sin{\theta^{e}_{12}}/ \sqrt{2}$. The approximate relation $\theta_{13} \approx \theta_{c} / \sqrt{2}$ is obtained if we take $\theta^{e}_{12} \approx \theta_{c}\approx 13^{0}$. The atmospheric and solar angles, $\sin{\theta_{12}}=1/\sqrt{3}$ and $\sin{\theta_{23}}=1/\sqrt{2}$ are in agreement with the TBC ansatz. Since TBC ansatz satisfies the experimentally observed angles, it  motivates the study of mixing in the charged lepton sector. 

The mixing in the first two generations of charged leptons are, in general, strongly constrained from charge flavour violation decays of mesons and muon. To illustrate this, consider a scalar operator in the mass basis, given by,
\begin{equation}
    \mathcal{O}^{S}_{mnij}=\frac{1}{\Lambda^{2}}J^{Q}_{mn}J^{L}_{ij},
\end{equation}
where,
\begin{align}
    J^{Q}_{mn}&=\mathcal{C}^{Q}_{mn}\bar{q}_{m}P_{R}q_{n} \nonumber\\
    J^{L}_{ij}&=\mathcal{C}^{L}_{ij}\bar{\ell}_{i}P_{R}\ell_{j}
\end{align}
Here, $\Lambda$ is the energy scale of the NP. In the above currents, $\ell$ represents a leptonic field and $q$ represents a quark field. $C^{Q}$ and $C^{L}$ are $3\times 3$ matrix, with $m,n$ representing quark generation and $i,j$ representing lepton generation. Rotating to the flavour basis,
\begin{align}\label{field relation}
    {\ell{'}_{L}}_{i} \rightarrow {(U^{L}_{e/\nu})}_{ik} {\ell_{L}}_{k};~~~ {q{'}_{L}}_{i} \rightarrow {(U^{L}_{q})}_{ik} {q_{L}}_{k} \ ,
\end{align}
where $\ell{'}$ and $q{'}$ represent the fermionic field in the flavour basis, the quark and lepton bilinears becomes,
\begin{align}
    J^{Q}_{mn}&=\mathcal{C'}^{Q}_{mn}\bar{q}{'}_{m}P_{R}q{'}_{n} \nonumber\\
    J^{L}_{ij}&=\mathcal{C'}^{L}_{ij}\bar{\ell}{'}_{i}P_{R}\ell_{j}{'}
\end{align}
In this article, since we aim to study the effects of mixing among charged leptons, we now focus only on the lepton sector. Using Eq. \eqref{field relation} the Wilson Coefficient in the flavour basis $\mathcal{C'}^{L}$ is related to the mass basis $\mathcal{C}^{L}$ by,
\begin{equation}
    \mathcal{C}^{L}_{ij} = \sum_{k}\mathcal{C'}^{L}_{kj} ({U^{L}_{e/\nu}})^{\dagger}_{ik} \ .
\end{equation}
Scalar and dipole operators where the effects of this mixing are prominent are given in Table \ref{tab:1}.
\begin{table}[b]
    \centering
    \begin{tabular}{|c|c|}
    \hline
         Scalar & Dipole \\
         \hline
          $\mathcal{O}^{SRR}_{\ell \ell q q}= ( \bar{\ell}_{L}~ \ell_{R})(\bar{q}_{L} ~ q_{R})$ & $\mathcal{O}^{TLR}_{\ell\ell } = ( \bar{\ell}_{L}\sigma^{\mu\nu}\ell_{R}) F_{\mu\nu}$\\
          $\mathcal{O}^{SRL}_{\ell\ell q q} = ( \bar{\ell}_{L}~ \ell_{R})(\bar{q}_{R} ~ q_{L})$ & $\mathcal{O}^{TLR}_{\nu \ell } = ( \bar{\nu}_{L}\sigma^{\mu\nu} \ell_{R})F_{\mu\nu}$\\
          $\mathcal{O}^{SRR}_{\nu \ell q q} = ( \bar{\nu}_{L}~ \ell_{R})(\bar{q}_{L} ~ q_{R})$&\\
          $\mathcal{O}^{SRL}_{\nu \ell q q} = ( \bar{\nu}_{L}~ \ell_{R})(\bar{q}_{R} ~ q_{L})$&\\
          \hline
    \end{tabular}
    \caption{Scalar and dipole operators in which we can realize the TBC mixing, where $\ell=\{e,\mu\}, \nu=\{\nu_{e},\nu_{\mu},\nu_{\tau}\}$ and $q=\{u,c,t\}/\{d,s,b\}$}
    \label{tab:1}
\end{table}
The relation between Wilson coefficients in mass basis and flavour basis is,
\begin{align}
    \mathcal{C}^{SRR}_{ijqq} &= \sum_{k}\mathcal{C'}^{SRR}_{kjqq} ({U^{L}_{e/\nu}})^{\dagger}_{ik} \nonumber\\
    \mathcal{C}^{SRL}_{ijqq} &= \sum_{k}\mathcal{C'}^{SRL}_{kjqq} ({U^{L}_{e/\nu}})^{\dagger}_{ik}\nonumber\\
   \mathcal{C}^{TLR}_{ij} &= \sum_{k}\mathcal{C'}^{TLR}_{kj} ({U^{L}_{e/\nu}})^{\dagger}_{ik}
   \end{align}
Operators $\mathcal{O}^{SRR}_{\ell \ell q q}, \mathcal{O}^{SRL}_{\ell \ell q q}$ and $\mathcal{O}^{TLR}_{\ell \ell}$ are sensitive to $U^{L}_{e}$, where as operator $\mathcal{O}^{SRR}_{\nu \ell q q},\mathcal{O}^{SRL}_{\nu \ell q q}$ and $\mathcal{O}^{TRR}_{\nu \ell }$ are sensitive to $U^{L}_{\nu}$.
Hence, the constraints on the operators $\mathcal{O}^{SRR}_{\ell \ell q q}$, $\mathcal{O}^{SRL}_{\ell \ell q q} $ and  $\mathcal{O}^{TLR}_{\ell \ell}$ are best studied using flavour violating decays like $\pi^{0} \rightarrow e^{+}\mu^{-}, K_{L} \rightarrow \mu^{+}e^{-}, \mu \rightarrow e \gamma $ and muon conversion ($\mu N \rightarrow eN$). In fact, if we assume that ${U^{L}_{e}}$ is the source of LFV, current experimental bounds from $K_{L} \rightarrow \mu^{+} e^{-}$, $\pi^{0} \rightarrow e^{+}\mu^{-}$ and $\mu \rightarrow e \gamma$ rules out the TBC mixing. 

We consider only scalar and dipole operator here as they are more constraining than the vector and tensor operator. The reason is that, in lepton flavour violating decays of mesons, the Chiral PT matching of vector operators brings in momentum dependence ($\propto p^{\pi^0}_\mu F_0$), while the scalar operators are proportional to $B_0F_0$, where $p^{\pi^0}_\mu$, $F_0$ and $B_0$ are the pion momentum, pion decay constant and low energy constant given by $B_0 = \frac{m_{\pi^0}^2}{m_u+m_d} = 2.667 GeV$ respectively. A detailed discussion can be found in \cite{Thomas:2022gak} where the authors show how relaxed the limits on vector operator are in comparison to the scalar operators for $\pi^0 \to \mu^{+} e^{-}$ decay. 

\subsection{$K_{L} \rightarrow \mu^{+} e^{-}$}
The Low Energy Effective Field theory operators in Table \ref{tab:1} need to be matched with operators in chiral perturbation theory to get meson decays. We use the matching and decay rate expressions used in Ref.\cite{Gasser:1983yg,Gasser:1984gg,Thomas:2022gak}. Operators that can contribute to $K_{L} \rightarrow \mu^{+} \mu^{-}$ at tree-level are,
\begin{align}
    K_{L} \rightarrow \mu^{+}\mu^{-} : ~~~ &  \mathcal{C}^{SRR}_{\mu \mu ds} \mathcal{O}^{SRR}_{\mu \mu ds}, \  \mathcal{C}^{SRR}_{\mu \mu sd} \mathcal{O}^{SRR}_{\mu \mu sd}, \  \mathcal{C}^{SRL}_{\mu \mu ds} \mathcal{O}^{SRL}_{\mu \mu ds}, \ \mathcal{C}^{SRL}_{\mu \mu sd} \mathcal{O}^{SRL}_{\mu \mu sd} 
\end{align}
The decay width of $K_{L} \rightarrow \mu^{+} \mu^{-}$ in terms of the Wilson coefficient is then given by,
\begin{align}\label{kltomumu}
    \Gamma(K_{L} \rightarrow \mu^{+}\mu^{-}) &= \frac{{|p_{\mu}|}_{K_{L}\mu\mu}}{8\pi m_{K_{L}}^{2}}\frac{B_{0}^{2} F_{0}^{2}}{4\Lambda^{4}}\left|\left(\frac{\mathcal{C}^{SRR}_{\mu \mu ds}}{\alpha}-\frac{\mathcal{C}^{SRR}_{\mu \mu sd}}{\beta}-\frac{\mathcal{C}^{SRL}_{\mu \mu ds}}{\alpha}+\frac{\mathcal{C}^{SRL}_{\mu \mu sd}}{\beta}\right)\right|^{2} (m_{K_{L}}^{2}-(m_{\mu}+m_{\mu})^{2}) 
    %&= (8.8\pm 0.25)\times 10^{-26} \text{GeV}
\end{align}
%\begin{equation}\label{ktomumu}
   % \left|\left(\frac{\mathcal{C}^{SRR}_{\mu \mu ds}}{\alpha}+\frac{\mathcal{C}^{SRR}_{\mu \mu sd}}{\beta}-\frac{\mathcal{C}^{SRL}_{\mu \mu ds}}{\alpha}-\frac{\mathcal{C}^{SRL}_{\mu \mu sd}}{\beta}\right)\right|^{2} = (0.791\pm0.022) \times 10^{-21} \text{GeV}^{-4}
%\end{equation}

%In the above expression $F_{0}=92.1\times10^{-3}$GeV is the pion decay constant in chiral limit and $B_{0}=2.667$ GeV is the non-perturbative low energy constant.
Here, $\alpha=\frac{1+\epsilon}{\sqrt{1+|\epsilon|^{2}}}$ and $\beta=\frac{1-\epsilon}{\sqrt{1+|\epsilon|^{2}}}$ where, $\epsilon$ is the CP violation parameter in the kaon oscillation with the value $|\epsilon|=2.228\times 10^{-3}$ \cite{Workman:2022ynf}. Values of other constants which are used is given in Table \ref{tab:Inputvalues}. In Eq. \eqref{kltomumu} $p_{\mu}$ is given by,

\begin{equation}
    |p_{\mu}|_{xyz} = \frac{1}{2 m_{x}} (m_{x}^{4} + m_{y}^{4} + m_{z}^{4}- 2 m_{x}^{2}m_{y}^{2} - 2 m_{x}^{2}m_{z}^{2} - 2 m_{y}^{2}m_{z}^{2})^{1/2}
\end{equation}

\begin{table}[h!]
    \centering
    \begin{tabular}{|m{1cm}|c|m{1cm}|c|}
    \hline
         & in GeV & & in GeV \\
         \hline
         $F_{0}$ & $92.1\times10^{-3}$ & $B_{0}$ & $2.667$\\
         $m_{K_{L}}$ & $(497.611 \pm 0.013) \times 10^{-3}$ &$m_{\pi^{0}}$ &$(134.9768 \pm 0.0005) \times 10^{-3}$\\
         $m_{e}$ & $0.511 \times 10^{-3}$ & $m_{\mu}$ & $105.65 \times 10^{-3}$\\
         \hline
    \end{tabular}
    \caption{Various input values in GeV.}
    \label{tab:Inputvalues}
\end{table}
By introducing LFV using $U^{L}_{e}$, the Wilson coefficients due to mixing in charged lepton sector become,
\begin{align}
    \mathcal{C}^{SRR}_{e\mu qq} &= \mathcal{C'}^{SRR}_{\mu \mu qq} ({U^{L}_{e}})^{\dagger}_{e \mu} \nonumber \\
    \mathcal{C}^{SRL}_{e\mu qq} &= \mathcal{C'}^{SRL}_{\mu \mu qq} ({U^{L}_{e}})^{\dagger}_{e \mu}
\end{align}
Thus the  lepton flavour violating decay of $K_{L} \rightarrow \mu^{+}e^{-}$, generated by the off-diagonal component of the charged lepton mixing matrix, $({U^{L}_{e}})_{e \mu}$, is given by,

\begin{align}\label{DRkltoemu}
    \Gamma(K_{L} \rightarrow \mu^{+}e^{-}) &= \frac{{|p_{\mu}|}_{K_{L}\mu e}}{8\pi m_{K_{L}}^{2}}\frac{B_{0}^{2} F_{0}^{2}}{4\Lambda^{4}}\left|\left(\frac{\mathcal{C}^{SRR}_{e\mu  ds}}{\alpha}-\frac{\mathcal{C}^{SRR}_{e\mu sd}}{\beta}-\frac{\mathcal{C}^{SRL}_{e\mu  ds}}{\alpha}+\frac{\mathcal{C}^{SRL}_{e\mu  sd}}{\beta}\right)\right|^{2} (m_{K_{L}}^{2}-(m_{\mu}+m_{e})^{2}) \nonumber\\
    &=\frac{{|p_{\mu}|}_{K_{L}\mu e}}{8\pi m_{K_{L}}^{2}}\frac{B_{0}^{2} F_{0}^{2}}{4\Lambda^{4}}|({U^{L}_{e}})_{e \mu}|^{2}\left|\left(\frac{\mathcal{C}^{SRR}_{\mu \mu ds}}{\alpha}-\frac{\mathcal{C}^{SRR}_{\mu \mu sd}}{\beta}-\frac{\mathcal{C}^{SRL}_{\mu \mu ds}}{\alpha}+\frac{\mathcal{C}^{SRL}_{\mu \mu sd}}{\beta}\right)\right|^{2} (m_{K_{L}}^{2}-(m_{\mu}+m_{e})^{2}) 
   % &< (6.04)\times 10^{-29} \text{GeV}
\end{align}
We take the ratio of flavour conserving  $\Gamma(K_{L} \rightarrow \mu^{+}\mu^{-})$ and the flavour violating decays $\Gamma(K_{L} \rightarrow \mu^{+}e^{-})$ to cancel out the hadronic factors and is given as,
\begin{align}
    \frac{\Gamma(K_{L} \rightarrow \mu^{+}e^{-})}{\Gamma(K_{L} \rightarrow \mu^{+}\mu^{-})}&={|({U^{L}_{e}})_{e \mu}|^{2}}\frac{{|p_{\mu}|}_{K_{L}\mu e}(m_{K_{L}}^{2}-(m_{\mu}+m_{e})^{2})} {{|p_{\mu}|}_{K_{L}\mu\mu}(m_{K_{L}}^{2}-(m_{\mu}+m_{\mu})^{2})}\nonumber \\
    &= 6.14 \times 10^{-2}
    \label{eq:kmuetbc}
\end{align}
In the above equation $|({U^{L}_{e}})_{e\mu}|^{2}\approx 0.05$. Using the experimental bounds on BR$(K_{L} \rightarrow \mu^{+} \mu^{-})_{exp} = (6.84 \pm 0.11) \times 10^{-9}$ \cite{Workman:2022ynf} and BR$(K_{L} \rightarrow \mu^{+} e^{-})_{exp}<4.7\times 10^{-12}$ \cite{BNL:1998apv} we get,
\begin{align}
    \frac{\Gamma(K_{L} \rightarrow \mu^{+}e^{-})_{exp}}{\Gamma(K_{L} \rightarrow \mu^{+}\mu^{-})_{exp}}<(0.687 \pm 0.011) \times 10^{-3}
\end{align}

This experimental result is in contradiction with the TBC mixing induced ratio given in Eq.~\eqref{eq:kmuetbc}, thus ruling out the TBC ansatz for operators containing down sector quarks. For completeness, we will also discuss the flavour violating pion decay.

 \subsection{$\pi^{0} \rightarrow e^{+}\mu^{-}$}
The operators that contributes to the flavour diagonal pion decay $\pi^{0} \rightarrow e^{+}e^{-}$ are,

\begin{align}
    \pi^{0} \rightarrow e^{+}e^{-} : ~~~ &  \mathcal{C}^{SRR}_{eeuu} \mathcal{O}^{SRR}_{eeuu}, \  \mathcal{C}^{SRR}_{eedd} \mathcal{O}^{SRR}_{eedd}, \  \mathcal{C}^{SRL}_{eeuu} \mathcal{O}^{SRL}_{eeuu}, \  \mathcal{C}^{SRL}_{eedd} \mathcal{O}^{SRL}_{eedd} 
\end{align}
With these operators, the decay rate of $\pi^{0} \rightarrow e^{+} e^{-}$ can be derived as,
\begin{align}
    \Gamma(\pi^{0} \rightarrow e^{+}e^{-}) &= \frac{{|p_{\mu}|}_{\pi^{0}e e}}{8\pi m_{\pi}^{2}}\frac{B_{0}^{2} F_{0}^{2}}{4\Lambda^{4}}|(\mathcal{C}^{SRR}_{eeuu}-\mathcal{C}^{SRR}_{eedd}-\mathcal{C}^{SRL}_{eeuu}+\mathcal{C}^{SRL}_{eedd})|^{2} (m_{\pi}^{2}-(m_{e}+m_{e})^{2}) 
    %&= (5.04 \pm 0.25)\times 10^{-16} \text{GeV}
    \label{eq:piondecayee}
\end{align}

Introducing LFV through $U^{L}_{e}$, like in the case with kaons, here, the flavour off-diagonal Wilson coefficients become,
\begin{align}
    \mathcal{C}^{SRR}_{\mu eqq} &= \mathcal{C}^{SRR}_{eeqq} ({U^{L}_{e}})^{\dagger}_{\mu e} \nonumber \\
    \mathcal{C}^{SRL}_{\mu eqq} &= \mathcal{C}^{SRL}_{eeqq} ({U^{L}_{e}})^{\dagger}_{\mu e} 
\end{align}

Using this relation, the flavour violating decay rate of $\pi^{0} \rightarrow e^{+}\mu^{-}$ can be computed from Eq.\eqref{eq:piondecayee} as,
\begin{align}\label{DRpitoemu}
    \Gamma(\pi^{0} \rightarrow e^{+}\mu^{-}) &= \frac{{|p_{\mu}|}_{\pi^{0}\mu e}}{8\pi m_{\pi}^{2}}\frac{B_{0}^{2} F_{0}^{2}}{4\Lambda^{4}}|(\mathcal{C}^{SRR}_{\mu   euu}-\mathcal{C}^{SRR}_{\mu edd}-\mathcal{C}^{SRL}_{\mu euu}+\mathcal{C}^{SRL}_{\mu edd})|^{2} (m_{\pi}^{2}-(m_{\mu}+m_{e})^{2}) \nonumber\\
    &= \frac{{|p_{\mu}|}_{\pi^{0}\mu e}}{8\pi m_{\pi}^{2}}\frac{B_{0}^{2} F_{0}^{2}}{4\Lambda^{4}}|({U^{L}_{e}})_{\mu e}|^{2}|(\mathcal{C}^{SRR}_{eeuu}-\mathcal{C}^{SRR}_{eedd}-\mathcal{C}^{SRL}_{eeuu}+\mathcal{C}^{SRL}_{eedd})|^{2} (m_{\pi}^{2}-(m_{\mu}+m_{e})^{2}) 
    %&< 2.49\times 10^{-18} \text{GeV}
\end{align}

Taking $|({U^{L}_{e}})_{\mu e}|^{2}\approx 0.05$, the ratio of $\Gamma(\pi^{0} \rightarrow e^{+}e^{-})$ with respect to $({U^{L}_{e}})_{e \mu}$ induced $\Gamma(\pi^{0} \rightarrow e^{+}\mu^{-})$ becomes,
\begin{align}
    \frac{\Gamma(\pi^{0} \rightarrow e^{+}\mu^{-})}{\Gamma(\pi^{0} \rightarrow e^{+}e^{-})} &= |({U^{L}_{e}})_{\mu e}|^{2} \frac{{|p_{\mu}|}_{\pi^{0}e \mu}(m_{\pi}^{2}-(m_{\mu}+m_{e})^{2}) }{{|p_{\mu}|}_{\pi^{0}e e}(m_{\pi}^{2}-(m_{e}+m_{e})^{2})} \nonumber\\
    & = 0.738 \times 10^{-2}
\end{align}
On the other hand, experimentally observed values for BR$(\pi^{0} \rightarrow e^{+}e^{-})_{exp}=( 6.46 \pm 0.33) \times 10^{-8}$ \cite{Workman:2022ynf} and BR$(\pi^{0} \rightarrow e^{+}\mu^{-})_{exp}< 3.2 \times 10^{-10}$ \cite{NA62:2021zxl}  give,
\begin{equation}
    \frac{\Gamma(\pi^{0} \rightarrow e^{+}\mu^{-})_{exp}}{\Gamma(\pi^{0} \rightarrow e^{+}e^{-})_{exp}} < (0.49 \pm 0.025) \times 10^{-2} \ .
\end{equation}

Though this is only slightly below the ratio predicted by TBC ansatz, future experiments with higher sensitivities will probe it better. 

\subsection{$ \mu \rightarrow e \gamma $}

Before computing the $\mu \to e \gamma$, lets first look at the magnetic and electric dipole moments of muon. The relevant Lagrangian term with MDM $a_{\ell}$ and EDM $d_{\ell}$
is given by,
\begin{equation}\label{dipole moment}
    \mathcal{L} = e  \frac{a_{\ell}}{4 m_{\ell}} \bar{\ell}\sigma^{\mu \nu}\ell F_{\mu \nu} - \frac{1}{2} i d_{\ell} \bar{\ell} \sigma^{\mu \nu} \gamma^{5} \ell F_{\mu \nu} \ .
\end{equation}
Comparing this with the tensor operator given in Table \ref{tab:1}, we get $a_{\ell} = 2 m_{\ell} \text{Re}(\mathcal{C}^{TLR}_{\ell \ell })$ and $d_{\ell} = \text{Im}\left(\mathcal{C}^{TLR}_{\ell \ell }\right)e$. Now, using the experimental results for magnetic and electric dipole moments of muon, $\Delta a_{\mu} =  a_\ell -a_{SM} = (25.1 \pm 5.9) \times 10^{-10}$ \cite{Muong-2:2021ojo} and $d_{\mu} < 1.9 \times 10^{-19} e~cm$ \cite{Muong-2:2008ebm}, these Wilson coefficients gets constrained as, $\text{Re}(\mathcal{C}^{TLR}_{\mu \mu }) = 1.19 \times 10^{-7} \text{GeV}^{-1}$ and $ \text{Im}\left(\mathcal{C}^{TLR}_{\mu \mu }\right) < 0.96 \times 10^{-5} \text{GeV}^{-1}$. 

Introducing the mixing matrix $U^{L}_{e}$, the  flavour violating Wilson coefficient becomes $\mathcal{C}^{TLR}_{e \mu } = \mathcal{C}^{TLR}_{\mu \mu } ({U^{L}_{e}})^{\dagger}_{e \mu}$. Thus the contribution to $\mu \rightarrow e \gamma$, is given by,
\begin{eqnarray}
    \text{Br}(\mu \rightarrow e \gamma) &=& \dis \frac{\tau_{\mu} \alpha m^{3}_{\mu}}{4} \left( |\mathcal{C}^{TLR}_{e \mu }|^{2} + |\mathcal{C}^{TRL}_{e \mu }|^{2}\right) \ , \nonumber \\
    &=& \dis \frac{2 \tau_{\mu} \alpha m^{3}_{\mu}}{4}|U^{L}_{e \mu}|^{2}  |\mathcal{C}^{TLR}_{\mu \mu }|^{2} \ ,
\end{eqnarray}
where $\tau_{\mu}=2.19 \times 10^{-6}s$ is the mean life time of muon. Computing the $\mu \to e \gamma$ using the Wilson Coefficient previously obtained, we get
\begin{align}
    \text{Br}(\mu \rightarrow e \gamma) &= 68.6  
\end{align}

The above value is much higher than the current upper bound $\text{BR}(\mu \rightarrow e \gamma)< 4.2 \times 10^{-13}$ \cite{MEG:2016leq}. 

Thus, the mixing matrix $U^{L}_{e}$, as given in Eq.~\eqref{eq:ule}, being the only source of charged lepton flavour violation is strongly disfavoured by the decays $K_{L} \rightarrow \mu^{+} e^{-}$, $\pi^{0} \rightarrow e^{+}\mu^{-}$ and $\mu \rightarrow e \gamma$ and rules out the Tri-bimaximal-Cabibbo mixing scenario in its original form. Whereas, we here show that Minimal Flavour Violation hypothesis become a natural framework in which they can exist.
% lepton flavour violating decays $K_{L} \rightarrow \mu^{+} e^{-}$, $\pi^{0} \rightarrow e^{+}\mu^{-}$ and $\mu \rightarrow e \gamma$ strongly constrain mixing of charged lepton sector. 
%One way to tackle this is using type I see-saw mechanism with partially constrained sequential right-handed neutrino dominance \cite{King:2012vj}. But this is only realized in models with addition $Z_{4}$ symmetry. In this article, we aim to show that the MFV hypothesis is a natural setting for TeV scale NP that mix the charged lepton sector.

\section{Minimal Flavour Violation with TBC mixing}
\label{sec:MFVtbc}
The Minimal Flavour Violation (MFV) hypothesis assumes that the Standard Model (SM) Yukawa couplings are the only source of flavour symmetry breaking \cite{DAmbrosio:2002vsn,Cirigliano:2005ck}. This means all the higher dimensional operators should be constructed out of the SM Yukawa couplings, satisfying the flavour symmetry $\mathcal{G}_{F}:SU(3)_{Q}\times SU(3)_{u}\times SU(3)_{d}\times SU(3)_{L}\times SU(3)_{e}$. The Yukawa couplings are considered as non-dynamical fields (spurions) which transform under the flavour symmetry $\mathcal{G}_{F}: \mathcal{G}_{QF} \times \mathcal{G}_{LF}$ ($\mathcal{G}_{QF}:SU(3)_{Q}\times SU(3)_{u}\times SU(3)_{d},~\mathcal{G}_{LF}:  SU(3)_{L}\times SU(3)_{e}$) as,
\begin{equation}
    Y_{u} \sim (3,\bar{3},1), ~~~~  Y_{d} \sim (3,1,\bar{3}), ~~~~ Y_{e} \sim (\bar{3},3) . \nonumber
\end{equation}
Higher dimensional operators are constructed using these Yukawa couplings satisfying the flavour symmetry $\mathcal{G}_{F}$. For the purpose of this article, we will assume MFV in the lepton sector and not in the quark sector. 

There are two different field contents possible for MFV hypothesis. First, with only SM fields called the minimal field content scenario and including the right-handed neutrinos called extended field content scenario \cite{Cirigliano:2005ck,Dinh:2017smk}. 

\subsection{Minimal Field Content}
In the case of minimal field content,  mass terms in the lepton sector are,
\begin{equation}
    \mathcal{L} = -vY^{ij}_{e} \bar{e}^{i}_{R} e ^{j}_{L} - \frac{v^{2}}{2\Lambda_{LN}}g^{ij}_{\nu} \overline{\nu^{ci}}_{L}\nu^{j}_{L} + h.c
\end{equation}
where, $\Lambda_{LN}$ is the scale of the lepton number violation and $v =174$ GeV is the vacuum expectation value of the Higgs field. The leptonic field transform under $\mathcal{G}_{LF}$ as:
\begin{equation}
    L_{L} \rightarrow V_{L}L_{L}, ~~~~  e_{R} \rightarrow V_{R}e_{R}
\end{equation}
In the above expression $L_{L}$ and $e_{R}$ represent $SU(2)$ doublet and singlet leptonic field. In order to keep the Lagrangian invariant under $\mathcal{G}_{LF}$, $Y_{e}$ and $g_{\nu}$ transform as:
\begin{equation}
    Y_{e} \rightarrow V_{R} Y_{e} V^{\dagger}_{L}, ~~~~ g_{\nu} \rightarrow V^{*}_{L} g_{\nu} V^{\dagger}_{L} 
\end{equation}
Assuming TBC mixing ansatz, the basis for MFV could be chosen as,
\begin{equation}
    Y_{e} = D_{e}{U^{L}_{e}}^{\dagger}, ~~~~ g_{\nu} = \frac{\Lambda_{LN}}{v^{2}} U^{L*}_{\nu}m_{\nu}U^{L\dagger}_{\nu}
\end{equation}
where $D_{e} = \frac{1}{v} diag(m_{e}, m_{\mu}, m_{\tau})$ and $m_{\nu}= diag(m_{\nu_{1}},m_{\nu_{2}},m_{\nu_{3}})$. Since this is different from the usual MFV where there is no mixing in the charged lepton sector, we call this scenario as Modified Minimal Flavour Violation (MMFV). A spurion that transforms as $(8,1)$ under the group $\mathcal{G}_{LF}$ can be constructed as $\Delta = g^{\dagger}_{\nu}g_{\nu}= \frac{\Lambda^{2}_{LN}}{v^{4}}U^{L}_{\nu}  m^{2}_{\nu} {U^{L}_{\nu}}^{\dagger}$.

\subsection{Extended Field Content}
On the other hand, introducing right-handed neutrino appends flavour symmetry in the lepton sector ($\mathcal{G}_{LF}$) to $\mathcal{G}^{'}_{LF} : \mathcal{G}_{LF} \times SU(3)_{\nu_{R}}$. Now, the mass term in the lepton sector for extended field content scenario becomes,
\begin{equation}
    \mathcal{L} = -vY^{ij}_{e} \bar{e}^{i}_{R} e ^{j}_{L} - vY^{ij}_{\nu} \bar{\nu}^{i}_{R} \nu^{j}_{L} - \frac{1}{2} M^{ij}_{\nu} \bar{\nu}^{ci}_{R}\nu^{j}_{R} + h.c
\end{equation}

The right-handed neutrino mass term breaks $SU(3)_{\nu_{R}}$ symmetry to $O(3)_{\nu_{R}}$ and they are assumed to be in their mass basis, that is $M^{ij}_{\nu}=M_{\nu}\delta^{ij}$. The Lagrangian remains invariant under the flavour symmetry $\mathcal{G}_{LF} \times O(3)_{\nu_{R}}$ if the field and spurions transform as, 
\begin{align}
      L_{L} \rightarrow &V_{L}L_{L}, ~~~~  e_{R} \rightarrow V_{R}e_{R}, ~~~~~ \nu_{R}\rightarrow O_{R}\nu_{R}\nonumber\\
    &Y_{e} \rightarrow V_{R} Y_{e} V^{\dagger}_{L}, ~~~~~~ Y_{\nu} \rightarrow O_{R} Y_{\nu} V^{\dagger}_{L}.
\end{align}
Generating an effective left-handed Majorana mass matrix by integrating out the right-handed neutrinos, we get,
\begin{equation}
    \frac{v^{2}}{\Lambda_{LFV}}g_{\nu} = \frac{v^{2}}{M_{\nu}}Y^{T}_{\nu}Y_{\nu}
\end{equation}
If we take $M_{\nu}=\Lambda_{LN}$ then, $g_{\nu} = Y^{T}_{\nu}Y_{\nu}$. Using $\mathcal{G}_{LF} \times O(3)_{\nu_{R}}$ symmetry, we rotate the fields such that there is mixing in the charged lepton sector. In this basis,
\begin{equation}
    Y_{e} = D_{e}{U^{L}_{e}}^{\dagger}, ~~~~Y^{T}_{\nu}Y_{\nu}= \frac{\Lambda_{LN}}{v^{2}} U^{L}_{\nu}m_{\nu}U^{L\dagger}_{\nu}
\end{equation}
and one can construct $\Delta = Y^{\dagger}_{\nu}Y_{\nu}$ such that it will transform as (8,1) under $\mathcal{G}_{LF}$. Since $U^{L}_{\nu}= U_{TB}$ is real, the CP violation in $U_{pmns}$ arises from the Cabibbo mixing matrix. As a result $Y^{\dagger}_{\nu}Y_{\nu}$ and $Y^{T}_{\nu}Y_{\nu}$ can be diagonalized by same unitary matrices. Then $\Delta = Y^{\dagger}_{\nu}Y_{\nu}=\frac{\Lambda_{LN}}{v^{2}}U^{L}_{\nu}  m_{\nu} {U^{L}_{\nu}}^{\dagger}$. Note that $\Delta$ is only proportional to $m_{\nu}$ whereas in minimal field content it is proportional to $m^{2}_{\nu}$.

\subsection{Analysis}
The operators that we consider are listed in Table \ref{tab:2}. Here we have kept only the dominant operators which are proportional to $\Delta$ and $Y_{e} \Delta$, and have neglected the operators that go as $Y_{e} Y^{\dagger}_{e}$ or higher orders of $Y_{e}$.
\begin{table}[h!]
    \centering
    \begin{tabular}{|c|c|c|}
    \hline
         Scalar &Dipole &Vector\\
         \hline
         $\mathcal{O}^{S1}= (\bar{L}_{L}\Delta^{\dagger}Y_{e}^{\dagger} e_{R})(\bar{d}_{R} \lambda^{S1}Q_{L})$& $\mathcal{O}^{T1} = H(\bar{L}_{L}\sigma^{\mu\nu}\Delta^{\dagger}Y_{e}^{\dagger} e_{R}) F_{\mu \nu}$ & $\mathcal{O}^{V1} = (\bar{L}_{L}\gamma^{\mu}\Delta L_{L})(\bar{Q}_{L}\lambda^{V1}\gamma_{\mu}Q_{L})$\\
         $\mathcal{O}^{S2} = -(\bar{L}_{L} \Delta^{\dagger}Y_{e}^{\dagger} e_{R})(\bar{Q}_{L} \lambda^{S2}i\tau^{2} u_{R})$& &$\mathcal{O}^{V2} = (\bar{L}_{L}\gamma^{\mu}\Delta L_{L})(\bar{L}_{L}\lambda^{V2}\gamma_{\mu}L_{L})$\\
         $\mathcal{O}^{S3}= (\bar{L}_{L}\Delta^{\dagger}Y_{e}^{\dagger} e_{R})(\bar{e}_{R} \lambda^{S3}L_{L})$&& $\mathcal{O}^{V1} = (\bar{L}_{L}\gamma^{\mu}\Delta L_{L})(\bar{u}_{R}\lambda^{V3}\gamma_{\mu}u_{R})$ \\
         && $\mathcal{O}^{V4} = (\bar{L}_{L}\gamma^{\mu}\Delta L_{L})(\bar{d}_{R}\lambda^{V4}\gamma_{\mu}d_{R})$\\
         \hline
         \end{tabular}
    \caption{ Operators satisfying flavour symmetry \cite{Cirigliano:2005ck}.  $Q_{L}$ and $L_{L}$ represents $SU(2)$ doublet quark and lepton. $u_{R}, d_{R}$ and $e_{R}$ represent $SU(2)$ singlet up quark, down quark and charged lepton respectively. $H$ is the SM Higgs field.}
    \label{tab:2}
\end{table}

These operators in minimal field content scenario and extended field content scenario with fermions in their mass basis are shown in Table \ref{tab:3a} and \ref{tab:3b}. Since we are interested in the protection offered by MFV hypothesis for charged lepton flavour violating currents induced by TBC mixing, we consider MFV only in the lepton sector and not in the quark sector\footnote{As the operators in Table \ref{tab:3a} and \ref{tab:3b} shows, if we assume Minimal Flavour Violation flavour symmetry, our results do not depend directly on $U^L_{e,\nu}$. Instead, they depend on $U_{pmns}$. On the other hand, lepton number violating processes can be shown to be proportional to $(U^{L}_e)^T U^L_\nu$. Since this is beyond the scope of this work we have not discussed it further in our work. Thus, new physics leading to mixing in the charged lepton sector finds a natural protection within MFV anzats. In addition to the lepton flavour violating currents discussed in \cite{Cirigliano:2005ck}, we also include $\mu \to eee$, $K^0 \to \mu e$ and $\pi^0 \to \mu e$. Note that, on matching with new physics, the dipole operator is generated at 1-loop, while the scalar operator (in Table \ref{tab:2}) generating $\mu \to eee$ is at tree-level. Thus, on translating the result to dynamical new physics parameters, $\mu \to eee$ will become important.}.

\begin{table}[h!]
    \centering
    \begin{tabular}{|c|c|}
    \hline
         & MMFV operators in minimal field content scenario\\
         \hline
         $\mathcal{O}^{S1}$ & $\frac{\Lambda^{2}_{LN}}{v^{4}}$ $\{(\bar{\nu}_{L} m^{2}_{\nu} U^{\dagger}_{pmns}D_{e} e_{R})(\bar{d}_{R}\lambda^{S1}u_{L})+ (\bar{e}_{L}  U_{pmns}  m^{2}_{\nu} U^{\dagger}_{pmns} D_{e} e_{R})(\bar{d}_{R} \lambda^{S1} d_{L})\}$\\
         $\mathcal{O}^{T1}$ & $\frac{\Lambda^{2}_{LN}}{v^{4}}$ $ (\bar{e}_{L} \sigma^{\mu\nu} v U_{pmns}  m^{2}_{\nu} U^{\dagger}_{pmns} D_{e} e_{R}) F_{\mu \nu}$\\
         $\mathcal{O}^{S2}$ &$\frac{\Lambda^{2}_{LN}}{v^{4}}$ $\{(\bar{\nu}_{L} m^{2}_{\nu} U^{\dagger}_{pmns}D_{e} e_{R})(\bar{d}_{L}\lambda^{S2} u_{R})-(\bar{e}_{L}  U_{pmns}  m^{2}_{\nu} U^{\dagger}_{pmns} D_{e} e_{R})(\bar{u}_{L} \lambda^{S2} u_{R})\}$\\
         $\mathcal{O}^{S3}$ & $\frac{\Lambda^{2}_{LN}}{v^{4}}$ $\{(\bar{\nu}_{L} m^{2}_{\nu} U^{\dagger}_{pmns}D_{e} e_{R})(\bar{e}_{R}\lambda^{S3}\nu_{L})+ (\bar{e}_{L}  U_{pmns}  m^{2}_{\nu} U^{\dagger}_{pmns} D_{e} e_{R})(\bar{e}_{R} \lambda^{S3} e_{L})\}$\\
         $\mathcal{O}^{V1}$& $\frac{\Lambda^{2}_{LN}}{v^{4}}$ $(\bar{\nu}_{L}\gamma^{\mu}m^{2}_{\nu} \nu_{L}+\bar{e}_{L}\gamma^{\mu}U_{pmns} m^{2}_{\nu} U^{\dagger}_{pmns} e_{L})(\bar{u}_{L}\lambda^{V1}\gamma_{\mu}u_{L}+\bar{d}_{L}\gamma_{\mu}\lambda^{V1}d_{L}) $\\
         $\mathcal{O}^{V2}$& $\frac{\Lambda^{2}_{LN}}{v^{4}}$ $(\bar{\nu}_{L}\gamma^{\mu}m^{2}_{\nu} \nu_{L}+\bar{e}_{L}\gamma^{\mu}U_{pmns} m^{2}_{\nu} U^{\dagger}_{pmns} e_{L})(\bar{\nu}_{L}\lambda^{V2}\gamma_{\mu}\nu_{L}+\bar{e}_{L}\lambda^{V2}\gamma_{\mu}e_{L})$\\
         $\mathcal{O}^{V3}$& $\frac{\Lambda^{2}_{LN}}{v^{4}}$ $(\bar{\nu}_{L}\gamma^{\mu}m^{2}_{\nu} \nu_{L}+\bar{e}_{L}\gamma^{\mu}U_{pmns} m^{2}_{\nu} U^{\dagger}_{pmns} e_{L})(\bar{u}_{R}\lambda^{V3}\gamma_{\mu}u_{R})$\\
         $\mathcal{O}^{V4}$& $\frac{\Lambda^{2}_{LN}}{v^{4}}$ $(\bar{\nu}_{L}\gamma^{\mu}m^{2}_{\nu} \nu_{L}+\bar{e}_{L}\gamma^{\mu}U_{pmns} m^{2}_{\nu} U^{\dagger}_{pmns} e_{L})(\bar{d}_{R}\lambda^{V4}\gamma_{\mu}d_{R})$\\
         \hline
    \end{tabular}
    \caption{MMFV operators in minimal field content scenario.}
    \label{tab:3a}
\end{table}

\begin{table}[h!]
    \centering
    \begin{tabular}{|c|c|}
    \hline
         & MMFV operators in extended field content scenario. \\
         \hline
         $\mathcal{O}^{S1}$ &$\frac{\Lambda_{LN}}{v^{2}}$ $\{(\bar{\nu}_{L} m_{\nu} U^{\dagger}_{pmns}D_{e} e_{R})(\bar{d}_{R}\lambda^{S1}u_{L})+ (\bar{e}_{L}  U_{pmns}  m_{\nu} U^{\dagger}_{pmns} D_{e} e_{R})(\bar{d}_{R} \lambda^{S1} d_{L})\}$\\
         $\mathcal{O}^{T1}$ &$\frac{\Lambda_{LN}}{v^{2}}$ $(\bar{e}_{L} \sigma^{\mu\nu}v U_{pmns}  m_{\nu} U^{\dagger}_{pmns} D_{e} e_{R}) F_{\mu \nu}$\\
         $\mathcal{O}^{S2}$ & $\frac{\Lambda_{LN}}{v^{2}}$ $\{(\bar{\nu}_{L} m_{\nu} U^{\dagger}_{pmns}D_{e} e_{R})(\bar{d}_{L}\lambda^{S2} u_{R})-(\bar{e}_{L}  U_{pmns}  m_{\nu} U^{\dagger}_{pmns} D_{e} e_{R})(\bar{u}_{L} \lambda^{S2} u_{R})\}$\\
         $\mathcal{O}^{S3}$ & $\frac{\Lambda_{LN}}{v^{2}}$ $\{(\bar{\nu}_{L} m_{\nu} U^{\dagger}_{pmns}D_{e} e_{R})(\bar{e}_{R}\lambda^{S3}\nu_{L})+ (\bar{e}_{L}  U_{pmns}  m_{\nu} U^{\dagger}_{pmns} D_{e} e_{R})(\bar{e}_{R} \lambda^{S3} e_{L})$\}\\
         $\mathcal{O}^{V1}$&$\frac{\Lambda_{LN}}{v^{2}}$  $(\bar{\nu}_{L}\gamma^{\mu}m_{\nu} \nu_{L}+\bar{e}_{L}\gamma^{\mu}U_{pmns} m_{\nu} U^{\dagger}_{pmns} e_{L})(\bar{u}_{L}\lambda^{V1}\gamma_{\mu}u_{L}+\bar{d}_{L}\gamma_{\mu}\lambda^{V1}d_{L}) $\\
         $\mathcal{O}^{V2}$&$\frac{\Lambda_{LN}}{v^{2}}$ $(\bar{\nu}_{L}\gamma^{\mu}m_{\nu} \nu_{L}+\bar{e}_{L}\gamma^{\mu}U_{pmns} m_{\nu} U^{\dagger}_{pmns} e_{L})(\bar{\nu}_{L}\lambda^{V2}\gamma_{\mu}\nu_{L}+\bar{e}_{L}\lambda^{V2}\gamma_{\mu}e_{L})$\\
         $\mathcal{O}^{V3}$& $\frac{\Lambda_{LN}}{v^{2}}$ $(\bar{\nu}_{L}\gamma^{\mu}m_{\nu} \nu_{L}+\bar{e}_{L}\gamma^{\mu}U_{pmns} m_{\nu} U^{\dagger}_{pmns} e_{L})(\bar{u}_{R}\lambda^{V3}\gamma_{\mu}u_{R})$\\
         $\mathcal{O}^{V4}$& $\frac{\Lambda_{LN}}{v^{2}}$ $(\bar{\nu}_{L}\gamma^{\mu}m_{\nu} \nu_{L}+\bar{e}_{L}\gamma^{\mu}U_{pmns} m_{\nu} U^{\dagger}_{pmns} e_{L})(\bar{d}_{R}\lambda^{V4}\gamma_{\mu}d_{R})$\\
         \hline
    \end{tabular}
    \caption{MMFV operators in extended field content scenario..}
    \label{tab:3b}
\end{table}
The effective Lagrangian that we consider in our analysis is of the form,
\begin{equation}
    \mathcal{L}=\frac{1}{\Lambda^{2}_{LFV}}\mathcal{C}\mathcal{O} \ ,
\end{equation}
where $\Lambda_{LFV}$ is the scale of NP that generates LFV. In our analysis, we take the lepton number breaking scale as $\Lambda_{LN}=10^{13}$ GeV \footnote{This is the upper limit value on $\Lambda_{LN}$ from perturbativity of $g_{\nu}$ and upper mass limit on light neutrino mass \cite{Cirigliano:2005ck}.} and $m_{\nu}=  diag\left(0,\sqrt{\Delta m^{2}_{12}},\sqrt{\Delta m^{2}_{12}+\Delta m^{2}_{23}}\right)$ for simplicity, with $\Delta m^{2}_{12}= 7.53 \times 10^{-5} \text{eV}^{2}$ and $\Delta m^{2}_{23}= 2.43 \times 10^{-3} \text{eV}^{2}$ \cite{Workman:2022ynf}. The mass of neutrinos is considered to be in normal hierarchy. There is no significant difference in the results if we consider inverted hierarchy. Below we discuss the limits on $\Lambda_{LFV}$ from various lepton flavour violating decays in both minimal field content scenario and extended field content scenario keeping the quark sector coupling ($\lambda$'s) arbitrary.\\

\begin{figure}[h]
\captionsetup[subfigure]{labelformat=empty}
    \centering
    \subfloat[(a) Minimal field content scenario]{\includegraphics[width=0.4\textwidth]{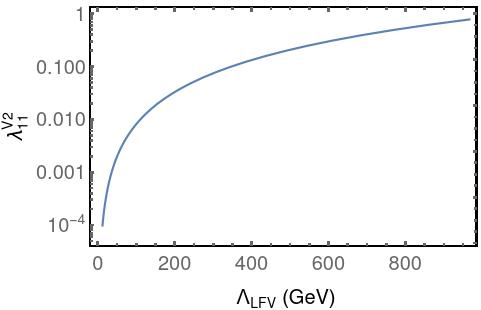}}\quad
    \subfloat[(b) Extended field content scenario]{\includegraphics[width=0.4\textwidth]{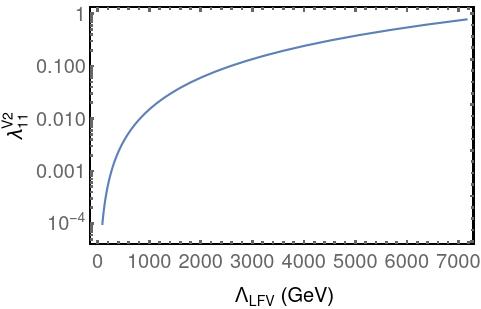}}
    \caption{The plots show limits on $\Lambda_{LFV}$ from BR$(\mu^{-}\rightarrow e^{-}e^{+}e^{-})_{exp}$ for different values of $\lambda^{V2}_{11}$ in both minimal and extended field content scenario. The region below the curve is allowed by the experiments}
    \label{fig:mutoeee}
\end{figure}

\noindent \underline{$\mu^{-}\rightarrow e^{-}e^{+}e^{-}:$}
\newline The operators that contribute to $\mu^{-}\rightarrow e^{-}e^{+}e^{-}$ are:
\begin{align}
   \mathcal{C}^{S3}_{e\mu e e} \mathcal{O}^{S3}_{e\mu e e} = &\Big(\bar{e}_{L} ~ \{\Delta^{\dagger}Y^{\dagger}_{e}\}_{12} ~\mu_{R}\Big)(\bar{e}_{R}~ \lambda^{S3}_{11} ~e_{L}) \nonumber\\
    \mathcal{C}^{V2}_{e\mu  e e}\mathcal{O}^{V2}_{e\mu  e e} = & \Big(\bar{e}_{L} ~ \gamma^{\mu}\Delta_{12} ~\mu_{L}\Big)(\bar{e}_{L}~ \gamma_{\mu}\lambda^{V2}_{11} ~e_{L})
\end{align}
The BR$(\mu^{-}\rightarrow e^{-}e^{+}e^{-})$ measured by SINDRUM collaboration gives an upper limit of $<1\times 10^{-12}$ \cite{SINDRUM:1987nra}. Branching ratio of $\mu^{-}\rightarrow e^{-}e^{+}e^{-}$ in case of a scalar operator \cite{Abada:2014kba} becomes,
\begin{align}
    &\text{BR}(\mu^{-}\rightarrow e^{-}e^{+}e^{-}) = \frac{1}{12}\frac{m_{\mu}^{5}\tau_{\mu}}{512\pi^{3}}\frac{|\mathcal{C}^{S3}_{e\mu ee}|^{2}}{\Lambda^{4}_{LFV}} \ , \nonumber\\
    &\frac{|\mathcal{C}^{S3}_{e\mu ee}|^{2}}{\Lambda^{4}_{LFV}}< 4.5 \times 10^{-21}\text{GeV}^{-4} \ .
\end{align}
Branching ratio in case of vector operator \cite{Abada:2014kba} is given by,
\begin{align}
    &\text{BR}(\mu^{-}\rightarrow e^{-}e^{+}e^{-}) = \frac{2}{3}\frac{m_{\mu}^{5}\tau_{\mu}}{512\pi^{3}}\frac{|\mathcal{C}^{V2}_{e\mu ee}|^{2}}{\Lambda^{4}_{LFV}} \ , \nonumber\\
    &\frac{|\mathcal{C}^{V2}_{e\mu ee}|^{2}}{\Lambda^{4}_{LFV}}< 5.6 \times 10^{-22} \text{GeV}^{-4} \ .
\end{align}
In the case of scalar operators, the constrain on the lepton flavour violation scale ($\Lambda_{LFV}$) is very weak due to an additional electron Yukawa coupling as given in Table. \ref{tab:2}.
%a TeV scale NP (ie. $\Lambda_{LFV} \approx $ TeV) gives a very low value for the Branching Ratio which can be probed in current and future experiments. 
Whereas, vector operators are not suppressed and Fig.~\ref{fig:mutoeee} shows the limits on $\Lambda_{LFV}$ from BR$(\mu^{-}\rightarrow e^{-}e^{+}e^{-})_{exp}$ for different values of $\lambda^{V2}_{11}$ in both minimal and extended field content scenarios. 
$\Lambda_{LFV}$ is more constrained in the case of extended field content scenario since $\Delta$ is proportional to the mass of the neutrino $m_{\nu}$ whereas, in minimal field scenario, it is proportional to the square of neutrino mass ($m^{2}_{\nu}$). \\

\noindent \underline{$\pi^{0} \rightarrow e^{+}\mu^{-}:$}
\newline The operators that contribute to $\pi^{0} \rightarrow e^{+}\mu^{-}$ are:
\begin{align}
    \mathcal{C}^{S1}_{\mu e d d}\mathcal{O}^{S1}_{\mu e d d} = &\Big(\bar{\mu}_{L} ~ \{\Delta^{\dagger}Y^{\dagger}_{e}\}_{21} ~e_{R}\Big)(\bar{d}_{R}~ \lambda^{S1}_{11} ~d_{L}) \nonumber\\
    \mathcal{C}^{S2}_{\mu e u u}\mathcal{O}^{S2}_{\mu e u u} = &\Big(\bar{\mu}_{L} ~ \{\Delta^{\dagger}Y^{\dagger}_{e}\}_{21} ~e_{R}\Big)(\bar{u}_{R}~ \lambda^{S2}_{11} ~u_{L}) \nonumber\\
    \mathcal{C}^{V1}_{\mu e q q}\mathcal{O}^{V1}_{\mu e q q} = & \Big(\bar{\mu}_{L} ~ \gamma^{\mu}\Delta_{21} ~e_{L}\Big)(\bar{u}_{L}~ \gamma_{\mu}\lambda^{V1}_{11} ~u_{L} + \bar{d}_{L}~ \gamma_{\mu}\lambda^{V1}_{11} ~d_{L}) \nonumber \\
     \mathcal{C}^{V3}_{\mu e u u}\mathcal{O}^{V3}_{\mu e u u } = & \Big(\bar{\mu}_{L} ~ \gamma^{\mu}\Delta_{21} ~e_{L}\Big)(\bar{u}_{R}~ \gamma_{\mu}\lambda^{V3}_{11} ~u_{R} ) \nonumber \\
    \mathcal{C}^{V4}_{\mu e d d}\mathcal{O}^{V4}_{\mu e d d} = & \Big(\bar{\mu}_{L} ~ \gamma^{\mu}\Delta_{21} ~e_{L}\Big)( \bar{d}_{R}~ \gamma_{\mu}\lambda^{V4}_{11} ~d_{R})
\end{align}

The branching ratio of $\pi^{0} \rightarrow e^{+}\mu^{-}$ in case of scalar operator, given in Eq.\eqref{DRpitoemu}, becomes,
\begin{align}
    &\text{BR}(\pi^{0} \rightarrow e^{+}\mu^{-}) = \frac{{|p_{\mu}|}_{\pi^{0}\mu e}}{8\pi m_{\pi}^{2}}\frac{\tau_{\pi^{0}}B_{0}^{2} F_{0}^{2}}{4}\frac{|\mathcal{C}^{S1}_{\mu e d d}+\mathcal{C}^{S2}_{\mu e u u}|^{2}}{\Lambda^{4}_{LFV}} (m_{\pi}^{2}-(m_{e}+m_{\mu})^{2}) \ , \nonumber\\
    &\frac{|\mathcal{C}^{S1}_{\mu e d d}+\mathcal{C}^{S2}_{\mu e u u}|^{2}}{\Lambda^{4}_{LFV}} <  4.1\times 10^{-13} \text{GeV}^{-4} \ ,
\end{align}
where $\tau_{\pi^{0}}=8.43 \times 10^{-17}s$ is the mean life time of pion. The branching ratio in terms of the Wilson coefficients of vector operator \cite{Thomas:2022gak} is given by,
\begin{align}
    &\text{BR}(\pi^{0} \rightarrow e^{+}\mu^{-}) = \frac{{|p_{\mu}|}_{\pi^{0}\mu e}}{8\pi m_{\pi}^{2}}\frac{ \tau_{\pi^{0}} F_{0}^{2}}{4}\frac{|\mathcal{C}^{V3}_{\mu e u u}-\mathcal{C}^{V4}_{\mu e d d}+\mathcal{C}^{V1}_{\mu e d d}-\mathcal{C}^{V1}_{\mu e u u}|^{2}}{\Lambda^{4}_{LFV}} (m_{\pi}^{2}(m_{\mu}^{2}+m_{e}^{2})-(m_{\mu}^{2}-m_{e}^{2})^{2}) \ , \nonumber \\
    & \frac{|\mathcal{C}^{V3}_{\mu e u u}-\mathcal{C}^{V4}_{\mu e d d}+\mathcal{C}^{V1}_{\mu e d d}-\mathcal{C}^{V1}_{\mu e u u}|^{2}}{\Lambda^{4}_{LFV}} <  2.67\times 10^{-10} \text{GeV}^{-4} \ .
\end{align}
We find that $\Lambda_{LFV}$ is weakly constrained by BR$(\pi^{0} \rightarrow e^{+}\mu^{-})_{exp}$ for both scalar operator and vector operator in minimal and extended field content scenarios.\\

\noindent \underline{$K_{L} \rightarrow \mu^{+}e^{-}:$}
\newline The operators that contribute to $K_{L} \rightarrow \mu^{+}e^{-}$ are:
\begin{align}
    \mathcal{C}^{S1}_{e\mu  d s}\mathcal{O}^{S1}_{e\mu  d s} = &\Big(\bar{e}_{L} ~ \{\Delta^{\dagger}Y^{\dagger}_{e}\}_{12} ~\mu_{R}\Big)(\bar{d}_{R}~ \lambda^{S1}_{12} ~s_{L}) \nonumber\\
    \mathcal{C}^{S1}_{e\mu  sd}\mathcal{O}^{S1}_{e\mu  sd} = &\Big(\bar{e}_{L} ~ \{\Delta^{\dagger}Y^{\dagger}_{e}\}_{12} ~\mu_{R}\Big)(\bar{s}_{R}~ \lambda^{S1}_{21} ~d_{L}) \nonumber\\
    \mathcal{C}^{V1}_{e\mu q q }\mathcal{O}^{V1}_{e\mu  q q} = & \Big(\bar{e}_{L} ~ \gamma^{\mu}\Delta_{12} ~\mu_{L}\Big)(\bar{d}_{L}~ \gamma_{\mu}\lambda^{V1}_{12} ~s_{L}+\bar{s}_{L}~ \gamma_{\mu}\lambda^{V1}_{21} ~d_{L}) \nonumber\\
    \mathcal{C}^{V4}_{e\mu q q}\mathcal{O}^{V1}_{e\mu  q q} = & \Big(\bar{e}_{L} ~ \gamma^{\mu}\Delta_{12} ~\mu_{L}\Big)(\bar{d}_{L}~ \gamma_{\mu}\lambda^{V4}_{12} ~s_{L}+\bar{s}_{L}~ \gamma_{\mu}\lambda^{V4}_{21} ~d_{L})
\end{align}

\noindent We assume that $\lambda^{S1}_{12}=\lambda^{S1}_{21} ,\lambda^{V1}_{12}=\lambda^{V1}_{21}$ and $ \lambda^{V4}_{12}=\lambda^{V4}_{21}$. The branching ratio in  case of scalar operator, given in Eq.\eqref{DRkltoemu}, now becomes,
\begin{align}
     &\text{BR}(K_{L} \rightarrow \mu^{+}e^{-}) = \frac{{|p_{\mu}|}_{K_{L}\mu e}}{8\pi m_{K^{0}_{L}}^{2}}\frac{\tau_{K_{L}} B_{0}^{2} F_{0}^{2}}{4}\left|\left(\frac{\mathcal{C}^{S1}_{e\mu  ds}}{\Lambda^{2}_{LFV}\alpha}-\frac{\mathcal{C}^{S1}_{e\mu  sd}}{\Lambda^{2}_{LFV}\beta}\right)\right|^{2} (m_{K^{0}_{L}}^{2}-(m_{\mu}+m_{e})^{2}) \ , \nonumber \\
     & \frac{|\mathcal{C}^{S1}_{e\mu d s}|^{2}}{\Lambda^{4}_{LFV}} <  2.2\times 10^{-20} \text{GeV}^{-4} \ ,
\end{align}
where $\tau_{K_{L}}=5.116 \times 10^{-8}s$ is the mean life time of $K_{L}$. And, the branching ratio in case of vector operator \cite{Thomas:2022gak} becomes,
\begin{align}\label{eq:kmesonvector}
    & \text{BR} (K_{L}\rightarrow e^{+}\mu^{-}) =  \frac{{|p_{\mu}|}_{K_L \mu e}}{8\pi m_{K^{0}_{L}}^{2}}\frac{ \tau_{K_{L}} F_{0}^{2}}{4}\left|\left(-\frac{\mathcal{C}^{V1}_{e\mu ds}}{\Lambda^{2}_{LFV}\alpha}+\frac{\mathcal{C}^{V1}_{e\mu  sd}}{\Lambda^{2}_{LFV}\beta}+\frac{\mathcal{C}^{V4}_{e\mu ds}}{\Lambda^{2}_{LFV}\alpha}-\frac{\mathcal{C}^{V4}_{e\mu  sd}}{\Lambda^{2}_{LFV}\beta}\right)\right|^{2} (m_{K^{0}_{L}}^{2}(m_{\mu}^{2}+m_{e}^{2})-(m_{\mu}^{2}-m_{e}^{2})^{2}) \ , \nonumber\\
    &  \left|\left(-\frac{\mathcal{C}^{V1}_{e\mu ds}}{\Lambda^{2}_{LFV}\alpha}+\frac{\mathcal{C}^{V1}_{e\mu  sd}}{\Lambda^{2}_{LFV}\beta}+\frac{\mathcal{C}^{V4}_{e\mu ds}}{\Lambda^{2}_{LFV}\alpha}-\frac{\mathcal{C}^{V4}_{e\mu  sd}}{\Lambda^{2}_{LFV}\beta}\right)\right|^{2} <  2.81\times 10^{-22} \text{GeV}^{-4} \ .
\end{align}
\begin{figure}[h]
\captionsetup[subfigure]{labelformat=empty}
    \centering
    \subfloat[(a) Vector operator minimal field content scenario]{\includegraphics[width=0.4\textwidth]{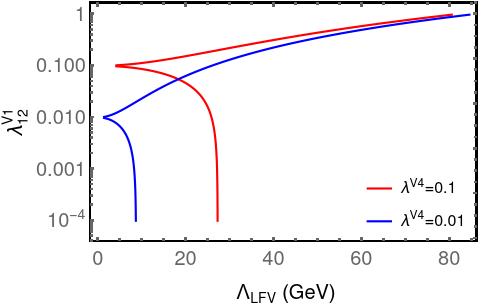}}\quad
    \subfloat[(b) Vector operator extended field content scenario]{\includegraphics[width=0.4\textwidth]{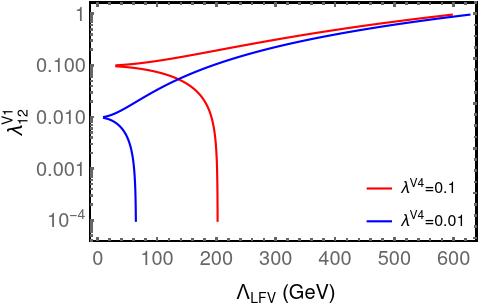}}\\
    \subfloat[(c) Vector operator minimal field content scenario]{\includegraphics[width=0.4\textwidth]{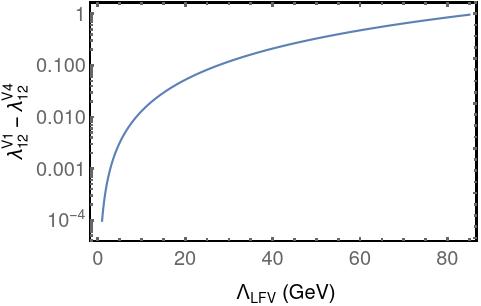}}\quad
    \subfloat[(d) Vector operator extended field content scenario]{\includegraphics[width=0.4\textwidth]{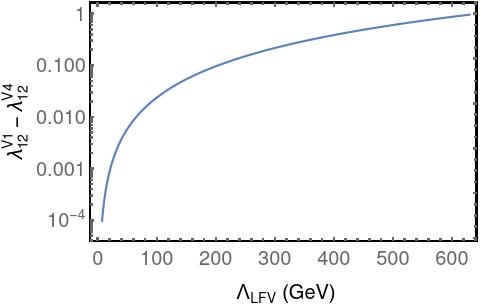}}
    \caption{The plots (a) \& (b) show limits on $\Lambda_{LFV}$ from $K_{L} \rightarrow \mu^{+}e^{-}$ treating $\lambda^{V1}_{12}$ as a variable while keeping $\lambda^{V4}_{12}$ fixed. While the plots (c) \& (d) shows the limit on $\Lambda_{LFV}$ keeping the combination $\lambda^{V1}_{12} - \lambda^{V4}_{12}$ as the variable. In the plots (a) \& (b), the red line corresponds to limits for benchmark value of $\lambda^{V4}_{12}=0.1$ and blue line corresponds to $\lambda^{V4}_{12}=0.01$. Note that there is a complete cancellation of Wilson coefficients, given in Eq.~\eqref{eq:kmesonvector}, for $\lambda^{V1}_{12}=\lambda^{V4}_{12}$, which leads to a beak like structure of the function.  Moreover, Eq.~\eqref{eq:kmesonvector} behaves differently for $\lambda^{V1}_{12}$ values above and below this point. While above this point $\mathcal{C}^{V1}$ is dominant and $\Lambda_{LFV}$ increases with $\lambda^{V1}_{12}$, below this point, $\mathcal{C}^{V4}$ becomes dominant. }
    \label{fig:kmesondecay}
\end{figure}

 The limit on $\Lambda_{LFV}$ from BR$(K_{L} \rightarrow \mu^{+}e^{-})_{exp}$ in vector operator scenario for different values of $\lambda$ is shown in Fig.\ref{fig:kmesondecay}. In the case of scalar operator $\Lambda_{LFV}$ is weakly constrained by the BR$(K_{L} \rightarrow \mu^{+}e^{-})_{exp}$.

\noindent \underline{$\mu N\rightarrow eN:$}
\newline The operators that contribute to $\mu N\rightarrow eN$ are:
\begin{align}
    \mathcal{C}^{S1}_{e \mu  d d} \mathcal{O}^{S1}_{e \mu d d} = &\Big(\bar{e}_{L} ~ \{\Delta^{\dagger}Y^{\dagger}_{e}\}_{12} ~\mu_{R}\Big)(\bar{d}_{R}~ \lambda^{S1}_{11} ~d_{L}) \nonumber\\
    \mathcal{C}^{S2}_{e \mu  u u}\mathcal{O}^{S2}_{e \mu u u} = &\Big(\bar{e}_{L} ~ \{\Delta^{\dagger}Y^{\dagger}_{e}\}_{12} ~\mu_{R}\Big)(\bar{u}_{R}~ \lambda^{S2}_{11} ~u_{L}) \nonumber\\
    \mathcal{C}^{V1}_{e \mu  q q}\mathcal{O}^{V1}_{e \mu  q q} = & \Big(\bar{e}_{L} ~ \gamma^{\mu}\Delta_{12} ~\mu_{L}\Big)(\bar{u}_{L}~ \gamma_{\mu}\lambda^{V1}_{11} ~u_{L} + \bar{d}_{L}~ \gamma_{\mu}\lambda^{V1}_{11} ~d_{L}) \nonumber \\
     \mathcal{C}^{V3}_{e \mu  u  u}\mathcal{O}^{V3}_{e \mu u u } = & \Big(\bar{e}_{L} ~ \gamma^{\mu}\Delta_{12} ~\mu_{L}\Big)(\bar{u}_{R}~ \gamma_{\mu}\lambda^{V3}_{11} ~u_{R}) \nonumber \\
      \mathcal{C}^{V4}_{e \mu d d}\mathcal{O}^{V4}_{e \mu d d} = & \Big(\bar{e}_{L} ~ \gamma^{\mu}\Delta_{12} ~\mu_{L}\Big)(\bar{d}_{R}~ \gamma_{\mu}\lambda^{V4}_{11} ~d_{R}) \nonumber \\
\end{align}
The best limit on $\mu \rightarrow e$ conversion comes from the SINDRUM II collaboration, which puts a limit on BR$(\mu Ti\rightarrow e Ti)_{exp}<6.1 \times 10^{-13}$ \cite{SINDRUMII:1998mwd}. The branching ratio of $\mu N\rightarrow eN$ conversion in the notation of Ref. \cite{Kitano:2002mt},
\begin{eqnarray}
    \text{BR}(\mu N\rightarrow e N)&=& \dis\frac{2G^{2}_{F}}{w_{capt}}\Bigg(|\tilde{g}^{(p)}_{LS}S^{(p)}+\tilde{g}^{(n)}_{LS}S^{(n)}+\tilde{g}^{(p)}_{LV}V^{(p)}+\tilde{g}^{(n)}_{LV}V^{(n)}|^{2}\nonumber\\
    & +& \dis |\tilde{g}^{(p)}_{RS}S^{(p)}+\tilde{g}^{(n)}_{RS}S^{(n)}+\tilde{g}^{(p)}_{RV}V^{(p)}+\tilde{g}^{(n)}_{RV}V^{(n)}|^{2}\Bigg)
\end{eqnarray}

where,
\begin{eqnarray}
        \tilde{g}_{LS, RS}^{(p)}
        & = &
        \sum_q G_{S}^{(q, p)} \  g_{LS, RS(q)}  \ ,
        \nonumber\\
        \tilde{g}_{LS, RS}^{(n)}
        & = &
        \sum_q G_{S}^{(q, n)} \  g_{LS, RS(q)}  \ ,
        \nonumber\\
        \tilde{g}_{LV, RV}^{(p)}
        & = &
        2 g_{LV, RV(u)} + g_{LV, RV(d)}\ ,
        \nonumber\\
        \tilde{g}_{LV, RV}^{(n)}
        & = &
        g_{LV,RV(u)} + 2 g_{LV, R(d)}\ .
\end{eqnarray}

\begin{figure}[t!]
\captionsetup[subfigure]{labelformat=empty}
    \centering
    \subfloat[(a) Scalar operator minimal field content scenario]{\includegraphics[width=0.4\textwidth]{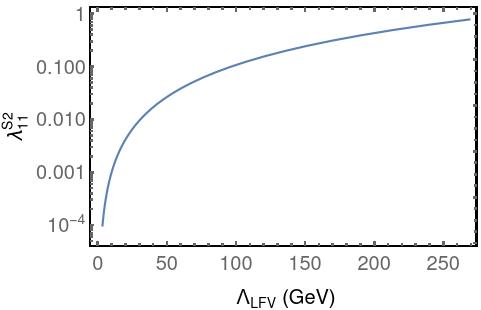}}\quad
    \subfloat[(b) Vector operator minimal field content scenario]{\includegraphics[width=0.4\textwidth]{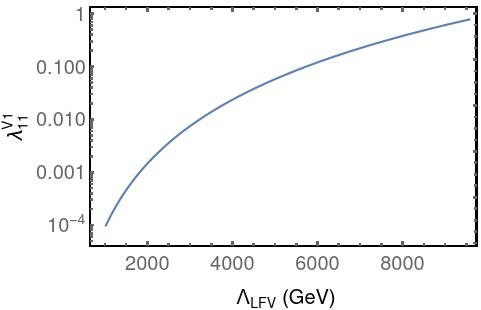}}\\
    \subfloat[(c) Scalar operator extended field content scenario]{\includegraphics[width=0.4\textwidth]{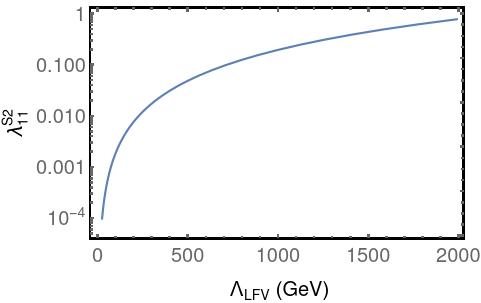}}\quad
    \subfloat[(d) Vector operator extended field content scenario]{\includegraphics[width=0.4\textwidth]{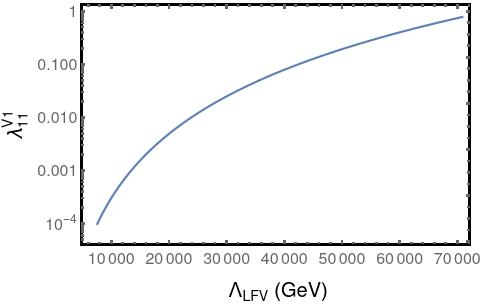}}
    \caption{The plots show limits on $\Lambda_{LFV}$ from BR$(\mu Ti\rightarrow e Ti)$ for different values of $\lambda^{S2}_{11}/\lambda^{V1}_{11}$ in scalar and vector operators in minimal and extended field content scenarios. The region below the curve is allowed by the experiments}
    \label{fig:mutoe}
\end{figure}

For scalar operator we consider a scenario when $\lambda^{S1}= 0$ and $\lambda^{S2}\ne0$, in this case $\mathcal{C}^{S2}_{e \mu  dd}=\frac{G_{F}}{2\sqrt{2}}g_{LS(d)}$,
\begin{align}\label{mutoeScalar}
    &\text{BR}(\mu Ti\rightarrow e Ti)=\frac{16}{w_{capt}}(G^{(d,p)}_{S}S^{(p)}+G^{(d,n)}_{S}S^{(n)})^{2}\frac{|\mathcal{C}^{S2}_{e \mu  dd}|^{2}}{\Lambda^{4}_{LFV}}\nonumber\\
     &\frac{|\mathcal{C}^{S2}_{e \mu  d d}|^{2}}{\Lambda^{4}_{LFV}} <  1.37\times 10^{-25} \text{GeV}^{-4}
\end{align}
In the case of a vector operator $\frac{G_{F}}{\sqrt{2}}g_{LV(d)}=(\mathcal{C}^{V1}_{e \mu dd}+\mathcal{C}^{V4}_{e \mu dd}), \frac{G_{F}}{\sqrt{2}}g_{LV(u)}=(\mathcal{C}^{V1}_{ e \mu uu}+\mathcal{C}^{V3}_{ e \mu uu})$,
\begin{align}\label{mutoeVector}
    &\text{BR}(\mu Ti\rightarrow e Ti)=\frac{36}{w_{capt}}(V^{(p)}+V^{(n)})^{2}\frac{|\mathcal{C}^{V1}_{e \mu dd}|^{2}}{\Lambda^{4}_{LFV}}\nonumber\\
    &\frac{|\mathcal{C}^{V1}_{e \mu  d d}|^{2}}{\Lambda^{4}_{LFV}} <  2.94\times 10^{-25} \text{GeV}^{-4}
\end{align}
Here we have assumed $\lambda^{V1}_{11}=\lambda^{V4}_{11}=\lambda^{V3}_{11}$ for simplicity. The constants that have used to compute the above expressions are listed in Table. \ref{tab:5}.

\begin{table}[h!]
    \centering
    \begin{tabular}{|c|c|c|c|}
    \hline
        $w_{cap}$ & $1.7047 \times 10^{18}$ GeV & $S^{(n)}$ & $0.0435~m^{5/2}_{\mu}$ \\
         $G^{(d,p)}_{S}$& $4.3$ & $V^{(p)}$ &$0.0396~ m^{5/2}_{\mu}$\\
         $G^{(d,n)}_{S}$ &$5.1$ & $V^{(n)}$ & $0.0468~m^{5/2}_{\mu}$\\
         $S^{(p)}$ &$0.0368~ m^{5/2}_{\mu}$ &&\\
          \hline
    \end{tabular}
    \caption{Various constants used in Eq. \eqref{mutoeScalar} and \eqref{mutoeVector} \cite{Kitano:2002mt}. }
    \label{tab:5}
\end{table}

 Fig.~\ref{fig:mutoe} shows the limit on $\Lambda_{LFV}$ from BR$(\mu Ti\rightarrow e Ti)$ for different values of $\lambda^{S2}_{11}/\lambda^{V1}_{11}$ in scalar and vector operator scenarios. We see that among all the lepton flavour violating decays considered, the best limit on $\Lambda_{LFV}$ arise from the vector operator. This is expected since vector operators are $\mathcal{O}(m_{\nu}^{2})$, were as the scalar operators are $\mathcal{O}(m_{\nu}^{2} m_{e})$.

\section{Conclusion}
\label{sec:conclusion}
Tri-bimaximal-Cabibbo mixing ansatz is a possible venue for new physics that can contribute towards charged lepton flavour violation. Importantly, this ansatz brings back the thought to be dead, but theoretically well-motivated, Tri-bimaximal mixing in the neutrino sector. In this paper, we investigate the LFV in charged lepton sector induced by TBC mixing in the context of various lepton flavour violating decays like $\mu \to eee$, $\mu \rightarrow e \gamma$,  $\pi^0 \to \mu e$, $K_L \to \mu e$ and $\mu N \to e N$. A model-independent analysis using the effective field theory operators revealed that if we assume TBC mixing to be the only source of LFV in charged lepton sector then it can be ruled out by current limits on lepton flavour violating decays $K_{L} \rightarrow \mu^{+} e^{-}$, $\pi^{0} \rightarrow e^{+}\mu^{-}$ and $\mu \rightarrow e \gamma$. 

Whereas, the MFV hypothesis protects the TBC mixing ansatz from these lepton flavour violating decays, naturally, making the NP available at LHC. We introduce MFV only in lepton sector and show that this suppresses the large flavour violations induced by the Cabibbo mixing in the first two generations of the charged leptons. In the MFV hypothesis, we have considered two scenarios. One with the minimal field content and the other including right-handed neutrinos namely, the extended field content. In the extended field content scenario, instead of introducing spurions, we have assumed the right-handed Majorana mass terms to be diagonal. With this, we have derived the limits on $\Lambda_{LFV}$ for both minimal and extended scenarios. In Table \ref{tab:MFContentSUM} and \ref{tab:EFContentSUM} we summarise the limit on $\Lambda_{LFV}$. The strongest constrain on $\Lambda_{LFV}$ arises from BR$(\mu Ti\rightarrow e Ti)$ and $\mu \to e \gamma$. Assuming $\mathcal{O}(1)$ couplings in the quark sector, this process bounds the cut-off scale to be $\Lambda_{LFV} \gtrsim 10$ TeV in the minimal content scenario. Instead, if we assume a slightly lower coupling of NP in the quark sector, this limit could be relaxed to $\Lambda_{LFV} \gtrsim 4$ TeV.
%In Table \ref{tab:MFContentSUM} and \ref{tab:EFContentSUM} we summarise the limit on $\Lambda_{LFV}$ for both the minimal field content and extended field content scenario in the lepton sector.
On the other hand, we find that the constraints in extended field content scenario are much stronger than the minimal field content scenario. The limit on $\Lambda_{LFV}$ is now $> 75$ TeV for $\mathcal{O}(1)$ coupling in the quark sector, while it relaxes to $31$ TeV on assuming more natural values of the couplings. This is expected since $\Delta$, given in Table. \ref{tab:2}, is proportional to $m_{\nu}$ in extended field scenarios whereas in minimal field scenario it is proportional to $m^{2}_{\nu}$. We also note that scalar operators are further suppressed compared to vector operators due to the presence of an additional charged lepton Yukawa coupling. 

Though the dipole operator is highly constraining, note that this operator arises at 1-loop in {\it Beyond Standard Model} (BSM). On the other hand, $\mu \to eee$ is a pure leptonic tree-level BSM process and hence, it is important that we compute them. With this consideration, we have found that in minimal field content scenario, $\mu \to eee$ constrains the new physics to $\Lambda_{LFV} \gtrsim 1$ TeV, while in extended field scenario, they become $\Lambda_{LFV} \gtrsim 8$ TeV.
  
Thus, assuming MFV, the lepton flavour violating new physics is not only closer to the scale of experiments, but also the Tri-bimaximal neutrino mixing matrix is safe from the reactor angle measurement.

\begin{table}[htb]
    \centering
    \begin{tabular}{|m{3cm}| c |c |c|c| }
    \hline
         Observables & Scenario & Limit on $\Lambda_{LFV}$ (TeV) &Scenario& Limit on $\Lambda_{LFV}$ (TeV)\\
         \hline
         \multirow{2}{12em}{\text{BR}$(\pi^{0} \rightarrow e^{+}\mu^{-})$}  & $\lambda^{S1}_{11}=1 ~~ \lambda^{S2}_{11}=0.1$ & $1\times 10^{-5}$ &$\lambda^{S1}_{11}=\frac{m_{d}}{v} ~~ \lambda^{S2}_{11}=\frac{m_{u}}{v}$ &$4.2\times 10^{-8}$\\
         & $\lambda^{V3}_{11}=1 ~~ \lambda^{V4}_{11}=0.1$ &$1.2 \times 10^{-3}$ & $\lambda^{V3}_{11}=\frac{m_u}{v} ~~ \lambda^{V4}_{11}=\frac{m_d}{v}$ &$4.9 \times 10^{-6}$\\
         \hline
         \multirow{2}{12em}{BR$(\mu^{-}\rightarrow e^{-}e^{+}e^{-})$}  & $\lambda^{S3}_{11}=1$ & $1.58 \times 10^{-2}$ &$\lambda^{S3}_{11}=c^{e}_{12}$ &$1.56 \times 10^{-2}$\\
         & $\lambda^{V2}_{11}=1$ &1.083   & $\lambda^{V2}_{11}=c^{e}_{12}$ &1.069\\
         \hline
         \multirow{2}{12em}{\text{BR} $(K_{L}\rightarrow \mu^{+}e^{-})$}  & $\lambda^{S1}_{12}=1$ & 0.01 &$\lambda^{S1}_{12}=\frac{m_{s}}{v}$ &$0.24 \times 10^{-3}$ \\
         & $\lambda^{V1}_{12}=1 ~~ \lambda^{V4}_{12}=0.1$ &0.274  & $\lambda^{V1}_{12}=\frac{m_{s}}{v} ~~ \lambda^{V4}_{12}=\frac{m_{s}}{v}$ &$8.6   \times 10^{-2}$\\
         \hline
         \multirow{2}{12em}{BR$(\mu Ti\rightarrow e Ti)$}& $\lambda^{S2}_{11}=1$ & 0.301 & $\lambda^{S2}_{11}=\frac{m_{d}}{v}$ & $1.6 \times 10^{-3}$\\
        & $\lambda^{V1}_{11}=1$ &10.126 & $\lambda^{V1}_{11}=\frac{m_{d}}{v}$ &0.728 \\
         \hline
         BR$(\mu \rightarrow e \gamma)$ & - & 4.17 &-&4.17\\
         \hline
         \end{tabular}
    \caption{Limit on $\Lambda_{LFV}$ from different lepton flavour violating decays in minimal field content scenario.}
    \label{tab:MFContentSUM}
\end{table}

\begin{table}[htb]
    \centering
    \begin{tabular}{|m{3cm}| c |c |c|c| }
    \hline
         Observables & Scenario & Limit on $\Lambda_{LFV}$ (TeV) &Scenario& Limit on $\Lambda_{LFV}$ (TeV)\\
         \hline
         \multirow{2}{12em}{\text{BR}$(\pi^{0} \rightarrow e^{+}\mu^{-})$}  & $\lambda^{S1}_{11}=1 ~~ \lambda^{S2}_{11}=0.1$ & $7.8\times 10^{-5}$ &$\lambda^{S1}_{11}=\frac{m_{d}}{v} ~~ \lambda^{S2}_{11}=\frac{m_{u}}{v}$  &$3.1\times 10^{-7}$\\
         & $\lambda^{S1}_{11}= 1 ~~ \lambda^{S2}_{11}=0.1$ &$9.21 \times 10^{-3}$ &$\lambda^{V3}_{11}=\frac{m_u}{v} ~~ \lambda^{V4}_{11}\frac{m_d}{v}$&$3.6\times 10^{-5}$\\
         \hline
         \multirow{2}{12em}{BR$(\mu^{-}\rightarrow e^{-}e^{+}e^{-})$}  & $\lambda^{S3}_{11}=1$ & 0.117 &$\lambda^{S3}_{11}=c^{e}_{12}$ &0.115\\
         & $\lambda^{V2}_{11}=1$ &8.019   & $\lambda^{V2}_{11}=c^{e}_{12}$ &7.915\\
         \hline
         \multirow{2}{12em}{\text{BR} $(K_{L}\rightarrow \mu^{+}e^{-})$}  & $\lambda^{S1}_{12}=1$ & 0.078 &$\lambda^{S1}_{12}=\frac{m_{s}}{v}$ & $1.8 \times 10^{-3}$ \\
         & $\lambda^{V1}_{12}=1 ~~ \lambda^{V4}_{12}=0.1$ &2  & $\lambda^{V1}_{12}=\frac{m_{s}}{v} ~~ \lambda^{V4}_{12}=\frac{m_{s}}{v}$ &0.636\\
         \hline
         \multirow{2}{12em}{BR$(\mu Ti\rightarrow e Ti)$}& $\lambda^{S2}_{11}=1$ & 1.577 & $\lambda^{S2}_{11}=\frac{m_{d}}{v}$ & $8.1\times 10^{-3}$\\
        & $\lambda^{V1}_{11}=1$ &74.97 & $\lambda^{V1}_{11}=\frac{m_{d}}{v}$ &5.397 \\
        \hline
        BR$(\mu \rightarrow e \gamma)$ & - & 30.94 &-&30.94\\
         \hline
         \end{tabular}
    \caption{Limit on $\Lambda_{LFV}$ from different lepton flavour violating decays in extended field content scenario.}
    \label{tab:EFContentSUM}
\end{table}

\acknowledgments
M.T.A. acknowledges the financial support of DST through INSPIRE Faculty grant DST/INSPIRE/04/2019/002507.

\bibliography{reference}

\end{document}